\documentclass[prb,twocolumn,floatfix,superscriptaddress,showpacs,preprintnumbers,yfonts,eufrak,amsmath,amssymb,showkeys,epsfig]{revtex4} 
\usepackage{graphicx}

\newcommand{\be}{\begin{equation}}
\newcommand{\ee}{\end{equation}}
\newcommand{\bea}{\begin{eqnarray}}
\newcommand{\eea}{\end{eqnarray}}
\newcommand{\br}{{\bf r}}
\newcommand{\Le}{\left}
\newcommand{\Ri}{\right}

\newcommand{\f}{\frac}

\begin{document}


\title{Elastic properties of graphene flakes: boundary effects and lattice vibrations}

\author{S. Bera}%
\affiliation{\mbox{Institut f\"ur Theorie der Kondensierten Materie, Karlsruhe Institute of Technology, D-76128 Karlsruhe, Germany}} 
\affiliation{Institut f\"ur Nanotechnologie, Karlsruhe Institute of Technology, D-76021 Karlsruhe, Germany}

\author{A. Arnold}%
\affiliation{Institut f\"ur Nanotechnologie, Karlsruhe Institute of Technology, D-76021 Karlsruhe, Germany}

\author{F. Evers}%
\affiliation{\mbox{Institut f\"ur Theorie der Kondensierten Materie, Karlsruhe Institute of Technology, D-76128 Karlsruhe, Germany}} 
\affiliation{Institut f\"ur Nanotechnologie, Karlsruhe Institute of Technology, D-76021 Karlsruhe, Germany}

\author{R. Narayanan}%
\affiliation{\mbox{Department of Physics, Indian Institute of Technology
    Madras, Chennai-600036, India}} 
\author{P. W\"olfle}%
\affiliation{\mbox{Institut f\"ur Theorie der Kondensierten Materie, Karlsruhe Institute of Technology, D-76128 Karlsruhe, Germany}} 
\affiliation{Institut f\"ur Nanotechnologie, Karlsruhe Institute of Technology, D-76021 Karlsruhe, Germany}

\date{\today}
\begin{abstract}
We present a calculation of the free energy, the surface free energy
and the  elastic constants (``Lam\'e parameters'' i.e, Poisson ratio, Young's 
modulus) of graphene flakes 
on the level of the density functional theory
employing different standard functionals.
We observe that the Lam\'e parameters in small
flakes can differ from the bulk values by 30\% for hydrogenated
zig-zag edges. The change results from the edge of the flake that
compresses the interior.
When including the vibrational zero point motion,
we detect a decrease in the bending rigidity, $\kappa$, 
by $\sim 26\%$. This correction is depending on the flake size, $N$,
because the vibrational frequencies flow with  growing $N$ 
due to the release of the edge induced compression. We calculate 
 Gr\"uneisen parameters and find good agreement with
previous authors. 
\end{abstract}



\maketitle


\section{Introduction}
Since its fabrication has become technologically
feasible\cite{discovery}, graphene has been in the
focus of frontier research\cite{kim05,Novo06,rise07}.
One of its most celebrated properties
are its massless low energy excitations\cite{Novo05,NetoRMP}
(``Dirac fermions''),
which emanate from the symmetries of the honeycomb lattice.
The electronic properties of graphene {\em flakes} are quite different from bulk
graphene due to the finite size and the presence of 
edges\cite{nakada96,Ihn09}. In particular, calculations suggest that
the zigzag edges of graphene nano-ribbons (quasi $1d$) have two flat bands
at the Fermi energy\cite{fujita96,brey06} that introduce
magnetism\cite{fujita98,louie06,sonPRL06,hod07,Kan08,peres06,pisani07,jiang07}. 
Recent theoretical studies on zigzag edged graphene flakes also confirm a tendency towards 
edge magnetism\cite{palaciosPRL,olegPRL,pisani07,jiang07}. 
This gives additional motivation for 
fabricating graphene based nano structures\cite{kern08,Danneau08,weber08,Xiaolin08,haluka07,Miao07}
for further studies.
The fabrication of such structures with well defined
edges still poses a considerable technological challenge.
Therefore, only very few experiments with structures exhibiting
zigzag edges have been reported\cite{ferrari08,parga08,Tan09}; a detailed
investigation of the edge physics still needs to be done.

An increased interest in the {\em elastic}
properties of graphene has developed 
recently\cite{Fasolino07,gui08,atalaya08,vanderZande08,kern08,Lee-expnt,castroStrain09,son09ar,faso09a,gustavo09,stefano09,Doussal09,reddy09,arroyo09,Ribeiro09}.
This is, for instance, because 
experiments suggest that graphene samples exhibit
a corrugated structure\cite{meyer07,stolyarova2007,ishigami07,parga08} (``ripples")
even at relatively low temperatures.
Their origin is thought to be due to residual
elastic strain produced by the experimental
preparation technique\cite{Lau-expnt}.

Also for elastic properties, edge effects can be highly relevant. 
For this reason, {\em flake} elastic properties are certainly interesting in their
own right. Namely, there is an intimate relation
between the electronic structure and the atomic geometry of graphene.
For example, the electronic spectrum of
a certain class of arm-chair graphene nano-ribbons is reported to
acquire a spectral gap due to an edge induced lattice dimerization along the
transport direction\cite{sonPRL06}.



\begin{figure}
\begin{center}
   \includegraphics[width=0.35\textwidth]{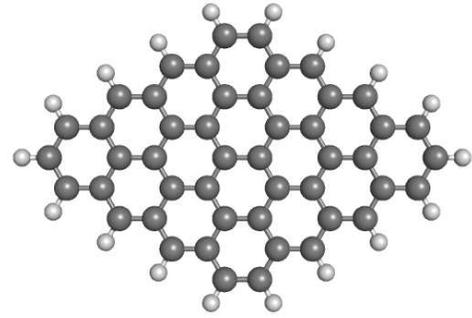}%
    \caption{Geometry of the $N{\times}N$ 
             hydrogenated graphene flakes(here $N{=}4$),
             that we have used for the 
             density functional calculations.}
\label{f1}
\end{center}
\end{figure}
In our study we investigate the free energy, the elastic properties,
and the phonon spectrum of $N\times N$ graphene sheets (``flakes'')
as displayed in Fig.~\ref{f1} using the density functional
theory (DFT).
We show, that the different chemical nature of C-C bonds at the
hydrogenated edge as compared to the bulk leads to an nearly
homogenous compression, i.e. strain.
As a consequence, the average C-C-distance in a $3\times 3$ flake
is reduced by a substantial amount, $0.3\%$;
for comparison, strain as achieved in typical pressure
experiments does not usually exceed values $\sim
1\%$\cite{Mohiuddin09, Lee-expnt,chenNL09}.

The presence of the surface induced strain leaves various traces in
the flakes' interior observables. (a) The flakes elastic constants, i.e. the
Lam\'e parameters, are enhanced as compared to the bulk case.
For isotropic strain in smallest flakes ($3\times 3$),
the (inverse) compressibility $\mu+\lambda$
(precise definition see below, Eq. (\ref{e2}))
increases by 30\%; for
shear forces the increase is even bigger, almost
a factor of 2. (b) Under bond compression the interatomic forces
typically increase, so that even the short wavelength vibrations, in
particular the optical phonons, exhibit a ``blue shift'' of their frequencies
with decreasing flake sizes.
This flow can be seen in the variation of the Raman spectra with strain
\cite{ferrari06,Shen08,ferrari08,Mohiuddin09,proctor09,tsoukleri09}
 and can be described in the standard
manner by Gr\"uneisen parameters.
 The values that we find here of Gr\"uneisen parameters,
agree reasonably well with previous reports\cite{Mohiuddin09,reich00,Shen08}.

Even though one might suspect, that our topics
have already been dealt with extensively in the
literature\cite{portal99,reich00,reich02,hod07a,tsoukleri09}, 
a detailed investigation
of the elastic properties of nano-flakes
is yet to be done; this refers, in particular, to an analysis of edge
and finite size effects of hydrogenated zig-zag flakes
which we perform in this work.

\section{Free energy of homogenous, planar flakes}
We consider the  free energy of a graphene flake as depicted in Fig.~\ref{f1}
with a homogenous C-C-distance, $d$, as the sum of all its bond energies:
\be
\label{e1}
 F_{N}(d) = N_{\rm i}\Psi(d) + N_{\rm e}\psi(d) + N_{\rm c}\psi^{\rm c}(d) +
 \frac{\phi^{\rm e}(d)}{N_{\rm e}} + \frac{\phi^{\rm i}(d)}{N_{\rm i}}
\ee
where $N_{\rm i}$ denotes the number of
internal C-C-bonds with an associated binding energy $\Psi$,
$N_{\rm e}$ denotes the number of edge located C-C-bonds
with energy $\psi$ 
and $\psi^{\rm c}$ includes the corner contributions, where $N_{\rm c}$ is the
 number of bonds linking the corner atoms,
 $N_{\rm c}{=}6$ in Fig.~\ref{f1}.
The binding energy per C-C-CH edge group
is close to $2\psi$ but not identical to it. For instance,
$\psi$ also includes
corrections of internal bonds, that still ``feel'' the presence
of the surface. Similarly, $N_c\psi^{\rm c}$ is approximating
the binding energy of the corner groups
(two HC-CH groups and two C-CH-C groups). 

We mention, that the representation (\ref{e1})
is slightly simplified in the following sense.
In general, the boundary (shape) of a given flake,
e.g as depicted in Fig.~\ref{f1},
does not share the hexagonal symmetry of the honeycomb lattice.
For this reason, in flakes with a fully relaxed atomic structure
bond lengths and bond angles are not strictly all the same.
Our DFT-calculations indicate, however,
that such distortions, though clearly detectable,
give only small corrections to those phenomenological parameters
that we are mostly interested in.

The {\em continuum} theory  of $2d$-membranes
has been devised for an inhomogeneous flake
with neighboring bonds exhibiting slowly (in space )
 varying bond distances, $d({\bf r})$,
and angles. In this formulation
the elastic energy is represented
by the functional\cite{lan00},
\bea
\label{e2}
	E&=&\f{\kappa}{2}
	\int_{\cal A} \!d^2{\bf r} (\nabla^2 h)^2 +
	\f{\mu+\lambda}{2}\int_{\cal A} d^2{\bf r} \  (u_{xx}{+}u_{yy})^2
	  \nonumber\\
        &&+ \f{\mu}{2}\int_{\cal
	  A}\!d^2{\bf r} \  \left[4  u_{xy}^2+ (u_{xx}{-}u_{yy})^2\right]. 
\eea
The flake coordinates are given with respect to a planar reference
state with area ${\cal A}$, that lives in the ${\bf r}{=}(x,y)$-plane; 
accordingly, the {\it in} plane coordinates constitute the displacement
vector, ${\bf u}(x,y)$, that measures the translation of each membrane
point $(x,y,z)$ with respect to the reference state.
The {\it out of} plane distortions define the height field $h(x,y)$;
for the planar case $h{=}0$. 
${\bf u}$ and $h$  together constitute the strain tensor ($i,j=x,y$),
\be\label{e3}
u_{ij}=\f{1}{2}\Le[\partial_iu_j+\partial_ju_i+\partial_ih\partial_jh
+\sum_k\partial_iu_k\partial_ju_k\Ri].
\ee
It is clear that the form of $u_{ij}$ depicted in Eq.~(\ref{e3}) is symmetric construction, with respect to the 
spatial derivatives of ${\bf u}$ ensure its invariance 
under in plane rotations by $90^\circ$ which correspond to 
${\bf u}\propto (y,-x)$.  

The total elastic energy (\ref{e2})
is a sum over contributions which resemble local oscillators
in the membrane plane.
The first term is proportionate to the curvature 
$\nabla^2 h$ and introduces the bending rigidity
$\kappa$.  It describes 
the energy cost for bending the membrane without changing the bond
lengths or in plane bond angles.  
\footnote{Note, that certain effects related to the thickness of the graphene
  sheet are left out in Eq.~(\ref{e2}). 
  For example, a linear displacement
  field $(h({\bf r}){=}x,{\bf u}{=}0)$ has no energy cost, because the
  description assumes, that such a conformation is equivalent to a
  rotation. However, this ignores that a rotation gives the
  $\pi_z$ orbitals a new direction in space, while the lifting up of
  atoms mediated by the linear displacement field does not. 
  In the latter case, there is an
  additional energy cost $\propto |\nabla h|^2$ 
  related to the fact, that the overlap of 
  $\pi_z$-orbitals changes, which is not included in
  Eq.~(\ref{e2}). 
Similarly, global rotational invariance is given only for free
flakes. It can be broken due to experimental boundary conditions,
e.g. the attachment of contacts. Also this may produce 
gradient terms $(\nabla h)^2$ in the functional (\ref{e2}).
}
The Lam\'e parameters, $\lambda$ and $\mu$, appearing in the second
and third term of Eq.~(\ref{e2}) describe the in plane rigidity. 

For homogenous, planar membranes
the elastic theory (\ref{e2})
may be considered as a continuum approximation to (\ref{e1})
which does not make explicit reference to
boundary terms. Edges are accounted 
for only in the boundary conditions
and (possibly) in a dependency
of the Lam\'e parameters on the
position with respect to the edge.
Usually not included in (\ref{e2}) is the fact that
this spatial dependency supports long range terms,
$\sim 1/{\rm flake size}$. They modify 
the Lam\'e parameters appearing in (\ref{e2})
even inside the flake's interior.

\subsection*{Phenomenological parameters}
\subparagraph*{Isotropic strain:}
In order to illustrate the cooperative effect between surface and
bulk, we consider an expansion of (\ref{e1}) 
in terms of the variable $\varepsilon{=}(d-d_0)/d_0$;
$\varepsilon$ quantifies the strain inside the flake.
The bulk free energy per bond has an expansion, 
\bea
\label{e4}
    \Psi &=& \Psi_0 + \frac{1}{2} \Psi_2 \ \varepsilon^2 
                    + \frac{1}{6} \Psi_3 \ \varepsilon^3 
                    + \frac{1}{24}\Psi_4 \ \varepsilon^4 \ldots 
\eea
where the bulk bond length $d_0$ is to 
be determined at $N_{{\rm i,e}}\to\infty$. 
The surface free energy may also be expanded about a minimum bond
length, $d_0^{\rm e}$, but in general $d_0^{\rm e}\neq d_0$.
After all, in the limit $N_{{\rm i,e}}\to\infty$,
just the first term in Eq.~(\ref{e1})
contributes to the free energy per area and therefore 
$d_0$ needs to minimizes $\Psi$, only.
Hence, we introduce the
relative deviation of surface and bulk optimal bond lengths,
${\frak d}{=}(d_0-d_0^{\rm e})/d_0$, so that we have an expansion
\bea
    \psi &=& \bar \psi_0  
    + \frac{1}{2} \bar \psi_2 \ (\varepsilon{+}{\frak d})^2  
    + \frac{1}{6} \bar \psi_3 \ (\varepsilon{+}{\frak d})^3 
    + \frac{1}{24}\bar \psi_4 \ (\varepsilon{+}{\frak d})^4 \ldots
    \nonumber \\
        &\equiv& \psi_0 + \psi_1 \ \varepsilon 
    + \frac{1}{2} \psi_2 \ \varepsilon^2  
    + \frac{1}{6} \psi_3 \ \varepsilon^3 + \ldots 
\label{e5}
\eea
where the coefficients in the second line are defined in terms of the
expansion the line before. 
The elastic properties of the flake are determined by the
expansion parameters $\Psi_{2,3,4}, \psi_{1,2,3}$.

At any finite value of $N_{{\rm i,e}}$,
optimization must also include the boundary (i.e. surface) terms and therefore
the optimal value of $\varepsilon$, $\varepsilon_{N}$, is non-vanishing
in this case; specifically,
\bea
\label{e6}
    \varepsilon_{N} &=&  -\frac{\psi_1}{\Psi_2} \frac{N_{\rm e}}{N_{\rm
        i}}\\
    &\approx& -{\frak d}\ \frac{\bar\psi_2}{\Psi_2}\frac{N_{\rm e}}{N_{\rm
        i}}.\nonumber   
\eea
In order to calculate the feedback of this shift into the
elastic parameters, we expand $F$ in the vicinity of its minimum,
$\varepsilon_{N}$, to the fourth order in $\varepsilon$. Recalling that
this corresponds to a strain ${\bf u}({\bf x}){=}\varepsilon {\bf x}$
we can compare the result with Eq.~(\ref{e2}) and thus find:
\bea
\label{e7}
    \mu+\lambda = \frac{1}{4}\left(\Psi_2
    + \frac{N_{\rm e}}{N_{\rm i}}\psi_2  \right)
+ \frac{1}{4}\Psi_3\varepsilon_{N}
      \nonumber\\
+ \frac{\varepsilon}{12}\left( \Psi_3
+ \frac{N_{\rm e}}{N_{\rm i}}\psi_3  + \Psi_4\varepsilon_{N} \right).   
\eea

The first  term in the rhs-expression
(\ref{e7}) simply accounts for the separate,
additive contributions of bulk and
surface (i.e. edge) free energies. 
The edge contribution, that appears here,
could formally be accounted for
in a generalized version of Eq.~(\ref{e2}) 
where one adds a boundary term.
Similarly, by allowing for a dependency of the Lam\'e
parameters on strain itself,
one could also include additive
an-harmonic effects, second bracket first two terms.
In either case, the phenomenological parameters are universal in the
sense, that they are the same for each flake size and geometry. 

The interesting pieces are the terms in $\varepsilon_{N}$, which
are mixing surface and bulk parameters:
$\Psi_3 \varepsilon_{N} \sim \psi_1\Psi_3/\Psi_2$. They encode
the ``cooperative'' effect between boundary induced
strain and bulk anharmonicities. It is due to them, that the flakes
elastic parameters need to be adjusted in principle
for every geometry separately.

\subparagraph*{Shear strain:}
An analogous analysis as for the isotropic strain also applies to shear
forces. The expansion is even in the shear strain ${\bf
  u}=(0,\varepsilon_{\rm s} x)$: 
\bea
\label{e8}
    \Psi &=& \Psi_0 + \frac{1}{2} \tilde \Psi_2 \ \varepsilon_{\rm s}^2 
                    + \frac{1}{24} \tilde \Psi_4 \ \varepsilon_{\rm s}^4 + \ldots \\
    \psi &=& \psi_0 + \frac{1}{2} \tilde \psi_2 \ \varepsilon_{\rm s}^2  
    + \frac{1}{24} \tilde \psi_4 \ \varepsilon_{\rm s}^4 + \ldots. 
\label{e9}
\eea
The new expansion parameters, $\tilde\Psi_{i},\tilde\psi_{i},
i=2,4,\ldots$, are, in general, dependent on the flake geometry. 
Again, by comparing to Eq.~(\ref{e2}) we find
\be
\mu = \tilde\Psi_2 + \tilde\psi_2 + \frac{1}{12} \varepsilon_{\rm s}^2
\left(\tilde\Psi_4 + \tilde \psi_4\right). 
\label{e10}
\ee
Here, surface and bulk free energies give strictly additive
contributions, and a cooperative effect does not emerge.

\section{Density functional calculations}

\subsection{Method} 
\label{abinitio} 

In Fig.~\ref{f1} we display the geometry of the $N\times
N$-graphene flake that is 
employed in our calculations: $N_{\rm i}=(N-1)(3N-1)$, $N_{\rm e}=8(N-1)$.
Electronic structure calculations
have been performed for a given
atomic configuration (C-C distance, flake geometry etc.)
on the basis of the density functional theory  as 
implemented in the quantum chemistry package
TURBOMOLE\cite{turbomole}.
We are comparing  GGA functionals (BP86\cite{BP86a,BP86b},
PBE\cite{pbe1,pbe2}) with a hybrid functional (B3LYP \cite{b3lyp})
and use a minimal basis set (SVP\cite{svpBasis}).
Specifically, we are working at zero temperature
and approximate the free (i. e. ground state) energy,
Eq.~(\ref{e1}),
by the DFT estimate for the total binding energy of the flake:
\be
\label{e11}
   F_{\rm el}(N,d) :=  E_{\rm el}(N,d) - E_{\rm free}(N). 
\ee
with,
\be
\label{e12}
    E_{\rm free}(N) = N_{\rm H} E_{\rm H} + N_{\rm C}E_{\rm C}
\ee
where $E_{\rm H/C}$ denote the DFT energies of a free charge neutral
hydrogen/carbon atom and $N_{\rm H/C}$ denotes the number of
hydrogen/carbon atoms in the flake.

\subsection{Results and Discussion}

\subparagraph*{Isotropic strain:}

\begin{figure}
\begin{center}
  \includegraphics[width=0.95\linewidth]{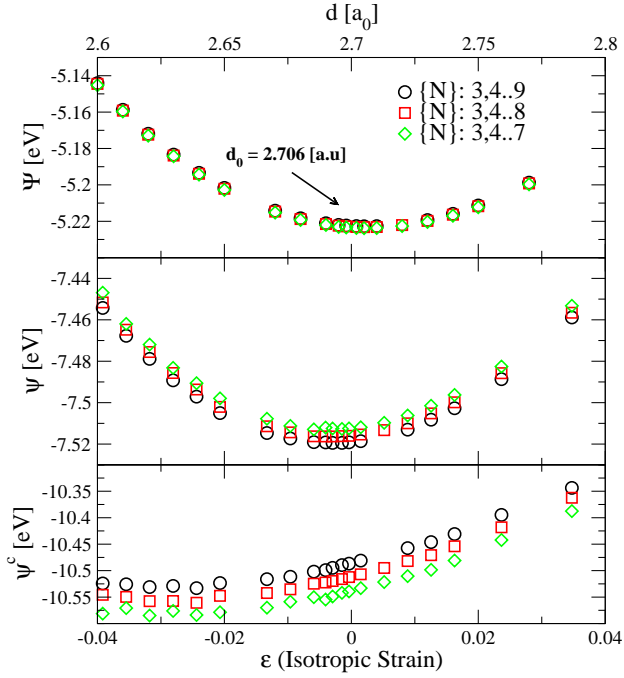}%
  \caption{Bulk ($\Psi$), surface (edge, $\psi$) and
    corner ($\psi^{\rm c}$) free energy
    per carbon bond in the graphene flakes, Fig.~\ref{f1},
    calculated with density functional theory
    (BP86-functional). Data based on the
     evaluation of three sets of flake sizes ranging from
     $\{N\}{=}3\ldots 9$, and a 5 parameter fit to Eqs.~(\ref{e1},\ref{e11})
     per $d$ values. ($a_0$:= 0.529 $\AA$.) }
  \label{f2}
\end{center}
\end{figure}

A sequence of DFT calculations has been performed
for $N{=}3\ldots 9$
and different values of the C-C distance, $d$.
For each distance, $F_{\rm el}(N,d)$ has been
calculated. In order to extract the expansion coefficients
of Eq.~(\ref{e1}),
$\Psi(d),\psi(d), \psi^{\rm c}(d), \phi^{\rm e}(d),
\phi^{\rm i}(d)$, we have performed five parameter fits on sets of 
raw DFT data. 
These fits were applied to three data sets consisting of
$\{N\}{=}3,\ldots 7$, $\{N\}{=}3\ldots 8$ and
$\{N\}{=}3\ldots 9$. 
The results for the surface, bulk and corner free energy have been
displayed in Fig.~\ref{f2}. The scatter between the fitting
parameters belonging to different data sets is relatively small,
which illustrates the stability of the fit.

The lattice constant of bulk graphene is estimated from the
minimum position of $\Psi(d)$ Fig.~\ref{f2}, upper panel 
 as $d_0{=}2.706a_0$,
where $a_0{=}0.529\AA$ denotes the Bohr radius. 
Comparing this position to the minimum of the
edge (surface) free energy, Fig.~\ref{f2}, center panel, 
$d_{0}^{\rm e}{=}2.694a_0$,
we find ${\frak d}{=}0.44$\%. This indicates
clearly the compression of the C-C bond length near the
edge. The shift of the minimum position 
to lower values 
becomes even more pronounced near
the corners, i.e. in $\psi^{\rm c}(d)$,
see Fig.~\ref{f2}, lower panel. 

Furthermore, we can perform a certain consistency check by
evaluating the bond energies. We have
a binding energy $\Psi(d_0){=}-5.22$eV for bulk carbon atoms.
The binding energy of $H$-atoms near edges (C-C-CH group)
is approximately
$\Delta E^{\rm e}_{\rm H}{\approx}
2\psi(d_0){-}2\Psi(d_0){\approx}-4.6$eV; when going to the corners
(HC-CH and C-CH-C groups) we have $\Delta E^{\rm c}_{\rm H}{\approx}
\psi^{\rm c}(d_0){-}\Psi(d_0)\approx -5.3$
which is roughly consistent with $\Delta E^{\rm e}_{\rm H}$, as it should be. 
\begin{table}
\begin{tabular}{c|c|c}
  & $d_0[\AA] $ &   $d_0^{\rm e}[\AA] $ \\\hline
{\small BP86} & $1.432^{\pm 0.002}$ & $1.427^{\pm0.001}$  \\\hline
{\small B3LYP}& $1.426^{\pm 0.001}$ & $1.421^{\pm0.002}$ \\\hline
{\small PBE}  & $1.431^{\pm 0.001}$ & $1.426^{\pm0.005}$ 

\end{tabular}
\caption{Minimum C-C bond length as extracted from bulk free energy
  and surface free energy correspondingly (see Fig.~\ref{f2}, upper and middle
  panel). The distance in
  $d_0$ and $d_o^{\rm e}$ leads to a nearly homogenous pressure on 
  the flake that modifies elastic and electronic-structure properties.
  Data is shown for three different functional used in DFT calculation.}
\label{t1}
\end{table}

To obtain also the other phenomenological parameters, a second
 (polynomial) fit of the traces $\Psi(d), \psi(d)$, Fig.~\ref{f2}, 
according to Eqs.~(\ref{e4},\ref{e5}) has been performed;
 all fitting parameters are summarized in Tab.~\ref{t1},\ref{t2} and \ref{t3}. 
\begin{table}
\begin{tabular}{c|c|c|c}
  & $-\Psi_0$\ [eV] & $\f{1}{2}\Psi_2$\ [eV] & $-\f{1}{6}\Psi_3$\ [eV]  \\\hline
  BP86   & $5.223^{\pm 0.001}$ & $46.301^{\pm
    0.082}$ & $128.301^{\pm 1.915}$ \\\hline
  B3LYP  & $5.008^{\pm 0.001}$  & $47.723^{\pm0.362}$ &
  $154.59^{\pm31.446}$ \\\hline
  PBE  & $5.373^{\pm 0.004}$ & $45.57^{\pm0.43}$
  & $187.309^{\pm24.448}$ 
\end{tabular}
  \caption{Bulk free energy coefficients as defined in Eq.~\eqref{e4}. 
  These coefficients are extracted from fitting Eq.~(\ref{e4}) to 
  the data in Fig.~\ref{f2}, upper panel.}
\label{t2}
\end{table}
\begin{table}
\begin{tabular}{c|c|c|c|c}
  & $-\psi_0$ \ [eV] & $\psi_1$\ [eV] & $\f{1}{2}\psi_2$\ [eV] & $-\f{1}{6}\psi_3$\ [eV]  \\\hline
  BP86    & $7.515^{\pm0.006}$ & $0.285^{\pm0.015}$ &
  $44.93^{\pm0.43}$ & $112.745^{\pm18.718}$     \\\hline
  B3LYP    & $7.187^{\pm0.006}$  & $0.261^{\pm0.015}$  &
  $45.586^{\pm2.154}$ & $ - $         \\\hline 
  PBE     & $7.63^{\pm0.01}$ & $0.308^{\pm0.011}$  & $47.503^{\pm1.082}$ & $-$ 
\end{tabular}
\caption{Edge (surface) coefficients as defined in Eq.~\eqref{e5}.
  These coefficients are extracted from fitting Eq.~\eqref{e5} to the 
  data in Fig.~\ref{f2}, middle panel.}
\label{t3}
\end{table}
\begin{figure}
  \begin{center}
    \includegraphics[width=0.85\linewidth]{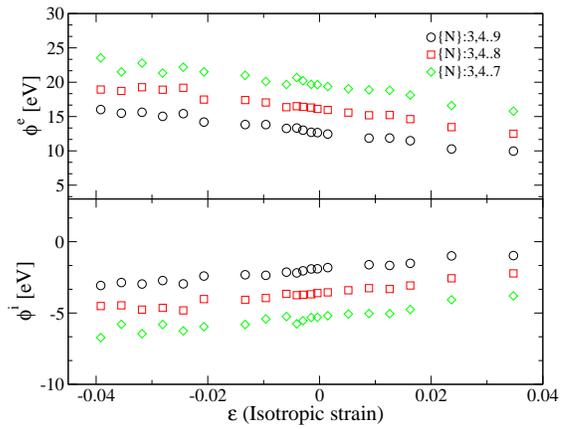}%
    \caption{Dependency of the amplitudes $\phi^{\rm i,e}$ describing
      the corrections in $1/N_{\rm e,i}$ to the binding energy
      $\Delta F_{\rm el}(N)$. Data were obtained by the 
      5 parameters fit, already underlying the traces shown
      in Fig.~\ref{f2}.}
    \label{f3}
  \end{center}
\end{figure}

When fitting the raw data to get $\Psi, \psi, \psi^{\rm c}$
the terms in $1/N_{\rm i,e}$
could not be neglected for the system sizes, that we considered.
The corresponding amplitudes are displayed in Fig.~\ref{f3}.
Unlike it was the case with the previous data, Fig.~\ref{f2},
the amplitudes $\phi^{\rm i,e}$ of the $1/N_{\rm i,e}$-corrections
still exhibit a considerable variation with increasing system size,
which is probably due to even higher order terms that have been
neglected in the expansion Eq.~(\ref{e1}).
Interestingly, while the magnitude of $\phi^{\rm i,e}(\varepsilon)$ is still
shifting the slope and perhaps also the sign of the two functions has
converged, already. Under this assumption we
may conclude, that both amplitudes flow closer
to zero values when $\varepsilon$ increases. This behavior is compatible
with the simple expectation, that the main effect incorporated in
the $1/N_{\rm i,e}$-corrections is the discreteness of the flake's
electronic spectrum with level spacings $\Delta_{\rm i,e}$ for bulk
and surface modes. With increasing $\varepsilon$ the bandwidth decreases
and so do $\Delta_{\rm i,e}$ and $\phi^{\rm i,e}$.

\begin{figure}
\begin{center}
  \includegraphics[width=0.85\linewidth]{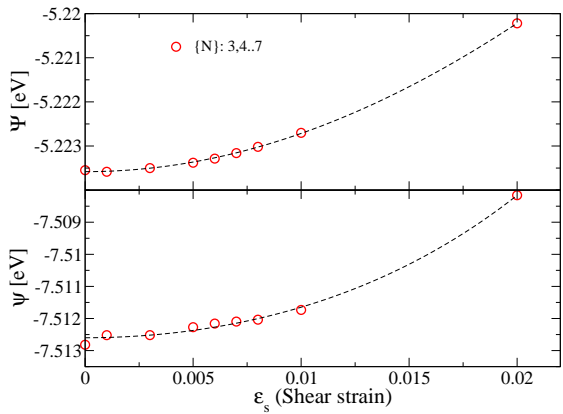}%
  \caption{Change in the free energy when applying a shear strain in 
    graphene flakes.
    Upper panel shows the change in bulk free energy per interior C-C bond; 
    lower
    panel exhibits change in surface (edge) free energy per C-C bond.
    Data sets are for system sizes \{N\}=3$\ldots$7 and extracted 
    from an expression analogous for shear
    to Eqs.~(\ref{e1},\ref{e11}). The lines indicate the polynomial fit
    according to Eqs.~(\ref{e8},\ref{e9}) with parameter given 
    in Tab.~\ref{t4}.} 
  \label{f4}
\end{center}
\end{figure}

\subparagraph*{Shear strain:}
A largely analogous method as was adopted for the
isotropic strain,
has also been applied for shear forces.
In this case, the convergence
of the DFT calculations turned out to be considerably more difficult, so
that the investigated system sizes range from $\{N\}{=}3\dots 7$, only.
From our fitting procedure we could determine the response of the bulk and
surface energy to the shear strain, $\varepsilon_{\rm s}$,
as shown in Fig.~\ref{f4}.
The parameters entering Eqs.~(\ref{e8},\ref{e9}) can be extracted and
are listed in Tab.~\ref{t4}.

\begin{table}
  \begin{tabular}{c|c|c|c|c}
    [eV]  & $\f{1}{2}\tilde \Psi_2$  & $-\f{1}{24}\tilde \Psi_4$ & $\f{1}{2}\tilde \psi_2$ & $\f{1}{24}\tilde \psi_4$ \\\hline 
    BP86    & 8.396$^{\pm0.426}$ & 871$^{\pm50\%}$ & 8.99$^{\pm3.29}$ &  $5237^{\pm60\%}$ 
  \end{tabular}
  \caption{Bulk and surface shear free energy coefficients as extracted
    from fitting Eq.~(\ref{e8}) (Eq.~(\ref{e9})) to the data in Fig.~\ref{f4} 
    upper panel (lower panel).}
  \label{t4}
\end{table}

\section{Flake elastic properties}
In the previous subsection, the focus was on the behavior of the free
energy on the flake size under isotropic and shear strain. In this
section, we discuss and illustrate what our previous findings imply for the
elastic properties of a single flake with a fixed size, $N$. Partly, we
are considering the same set of data again, but now plotting observables
directly for $N$ fixed. 

\subsubsection{Homogenous isotropic strain}
\begin{figure}
\begin{center}
   \includegraphics[width=0.95\linewidth]{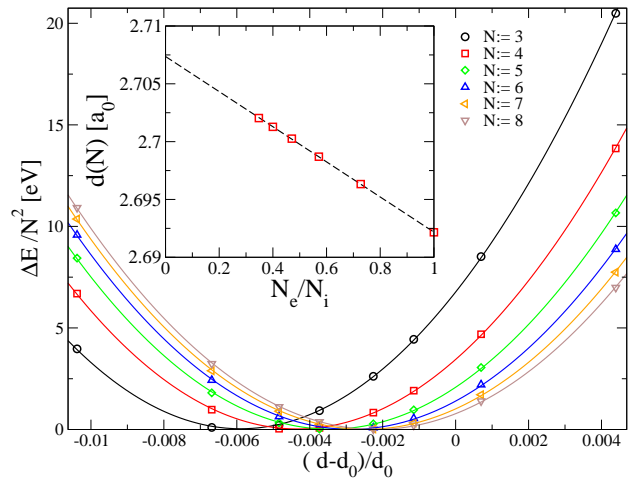}%
    \caption{Excess energy $\Delta E$ per unit cell generated
      by rescaling all bond length, $d$,
      (homogeneous, isotropic strain). $\Delta E$ exhibits flow of the
      equilibrium bond length with the linear flake size $N$.
      Main panel: $\Delta E$ over bond length $d$ employed in the
      simulation. $d$ is measured relative to the bulk bond length.
      The lines serve as a guide to the eyes.  
      Inset: Extrapolating the equilibrium
      bond length $d({N})$ into the bulk limit: $d_{\rm
        0}=2.707\pm 0.001$. }
\label{f5}
\end{center}
\end{figure}

Fig.~\ref{f5} shows, how the excess energy per unit cell grows under
increasing strain for different flake sizes $N$.
It is readily seen from this plot, that there is a
shift of the equilibrium lattice constant $d(N)$ to smaller values.
In the light of the previous section, this shift is the expected
consequence of the surface induced strain $\varepsilon_{N}$. The inset shows,
the scaling with $N_{\rm e}/N_{\rm i}$.

\begin{figure}
\begin{center}
   \includegraphics[width=0.95\linewidth]{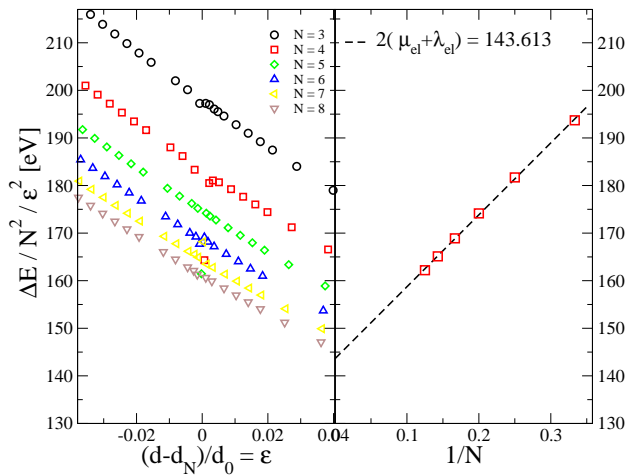}%
    \caption{Estimating the sum of
      the Lam\'e parameters $\mu_{\rm el}+\lambda_{\rm el}$ and the boundary
      correction (offset of traces) from
      $\Delta E$ displayed in the previous Fig.~\ref{f5}.
      Left: Data for curvature exhibit a slope which indicates the linear
      dependency of the Lam\'e parameter $\lambda_{\rm{el}}$ on strain.
      (Linear terms in $\mu_{\rm{el}}$ do not appear, see Fig.~\ref{f9}.)
      Right: extrapolating the curvature at $\varepsilon=0$
      into the bulk limit. }
\label{f6}
\end{center}
\end{figure}

In addition, we also extract the flake elastic constant $\mu{+}\lambda$
from the parabolic shape of the curves, Fig.~\ref{f5}. To this end, we replot
the data in Fig.~\ref{f6} left, so as to highlight the curvature and its
strain dependency. On the basis of Eq.~(\ref{e7}) we can conclude,
that the offset of the curves is a consequence of (a)
the presence of the surface and the extra energy required for its 
compression (term $N_{\rm e}/N_{\rm i}$ in Eq.~(\ref{e7}))
and (b) the feedback of the surface strain $\varepsilon_{N}$ 
into the bulk C-C distance. Extrapolating the zero strain values into
the limit, $N{\to}\infty$, we recover the previously derived bulk
limit value. This check is displayed in Fig.~\ref{f6}, right. The plot also
reveals, that the deviation of elastic parameters from their bulk
values in small flakes may not be very small. For our smallest flakes,
$N{=}3$ it reaches almost 30\%. 

Additional information can be extracted from Fig.~\ref{f6} left, about
anharmonicities which manifest themselves in the slope of the curves
displayed. This pre-factor of the an-harmonic term (linear in $\varepsilon$)
in Eq.~(\ref{e7}) admits the following interpretation.
The slope changes with increasing $N$ since the contributions of the
surface ($\psi_3$-term) and the surface induced bulk compression
($\Psi_4\varepsilon_{N}$-term) diminish. 
A non-vanishing value of $\Psi_3/12$ for the slope
will remain however even in the bulk limit. 

\subsubsection{Shear strain}
\begin{figure}
\begin{center}
   \includegraphics[width=0.95\linewidth]{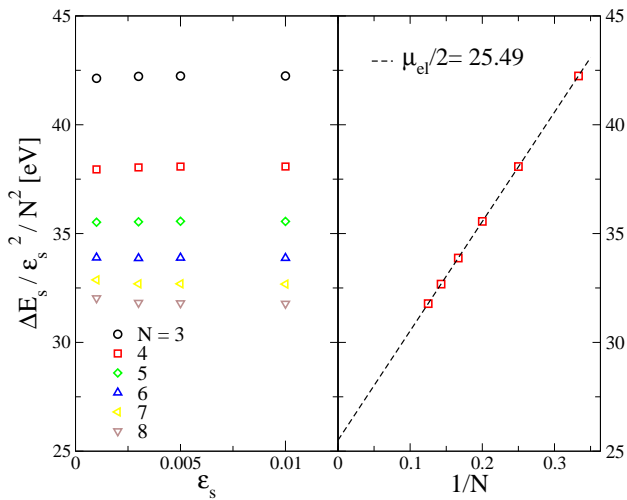}%
    \caption{Estimate for Lam\'e parameter $\mu_{\rm el}$ determined from the
      excess energy
      $\Delta E_{\rm{s}}$ per unit cell under pure shear strain with strength
      $\varepsilon_{\rm s}$. (Procedure similar to previous Fig.~\ref{f6}.)
      Left panel: dependency of curvature of 
      $\Delta E_{\rm{s}}(\varepsilon_{\rm{s}})$
      on the linear flake size $N$. Due to mirror symmetries of the
      unit cell, linear corrections do not appear for the shear
      parameter $\mu_{\rm{el}}$. Right: extrapolating the
      curvature into the bulk limit.}
\label{f7}
\end{center}
\end{figure}

Following the same strategy as we did before with Fig.~\ref{f6},
we plot in Fig.~\ref{f7} the excess energy $\Delta E_{\rm s}$
induced by pure shear strain,
${\bf u}({\bf x}) = \varepsilon_{\rm s}(0,x)$.
Again, the plot emphasizes
the curvature in this quantity, $\mu$,
and how it evolves with the flake size.
Since $\Delta E_{\rm s}$ is even in the shear strain, only
positive values of $\varepsilon_{\rm s}$ are given.  
Also, for the same reason an-harmonic terms exist only in the quartic
order, so that the displayed data traces have zero slope.
Similar to the previous case of
isotropic strain, we also witness here a very strong dependency of the
elastic constant on the flake size. In fact, for shear strain it
reaches almost 70\% for the small system sizes that we are considering. 

\begingroup
\squeezetable
\begin{table}
\begin{tabular}{c|c|c|c|c|c}
  & $d_{0}[\AA]$ & $\mu_{\rm{el}}+\lambda_{\rm{el}}$[eV] & $\mu_{\rm{el}}$[eV] & $\nu$ & Y[N$/$m]\\\hline
  BP86  & 1.432$^{\pm0.001}$  & 70.715$^{\pm0.011}$ & 50.95$^{\pm0.01}$ & 0.162 & 356.23 \\\hline
  B3LYP & 1.427$^{\pm0.001}$  & 71.21$^{\pm0.12}$   & - & - & - \\\hline 
  PBE   & 1.431$^{\pm0.002}$  & 69.027$^{\pm0.012}$ & - & - & -  \\ \hline
  prev. & 1.42[\onlinecite{faso09a,Marzari05}] & 66.571[\onlinecite{faso09a}]& 49.45[\onlinecite{faso09a}] & 0.149[\onlinecite{kudin01}],& 346[\onlinecite{faso09a}]\\
  calc. & 1.41[\onlinecite{Ribeiro09}]& & & 0.173[\onlinecite{gui08}] & 307[\onlinecite{reddy09}]
  \\ 
  & 1.45[\onlinecite{arroyo04}] & & & 0.16[\onlinecite{arroyo04}] & 345[\onlinecite{kudin01}] \\ 
& & & & 0.31[\onlinecite{stefano09}] & 336[\onlinecite{arroyo04}] \\
& & & & & 312[\onlinecite{stefano09}] \\ \hline
  expt. & - & - & - & - &  342[\onlinecite{Lee-expnt}] \\ 
  (Graphene) & & & &\\\hline
  expt. & 1.421[\onlinecite{hcp}] & - & - & 0.165[\onlinecite{blakslee70}] &  371[\onlinecite{krisch07}]\footnote{assuming graphene thickness 0.335 nm.}\\ 
  (Graphite) & 1.422[\onlinecite{krisch07}] & & & & \\ \hline
\end{tabular}
\caption{Comparison of C-C-bond distance in bulk graphene, elastic constants 
  , Poisson ratio($\nu$)  and Young's modulus(Y) as extracted from 
  Fig.~(\ref{f5},\ref{f6},\ref{f7}) respectively by extrapolating the values in bulk limit ($N \rightarrow \infty$) with previous works.
  Data is shown for three different functionals we
  used in DFT calculation.}
\label{t5}
\end{table}
\endgroup

\subsubsection{Buckling induced strain}
\begin{figure}[t]
\begin{center}
  \includegraphics[width=0.5\linewidth]{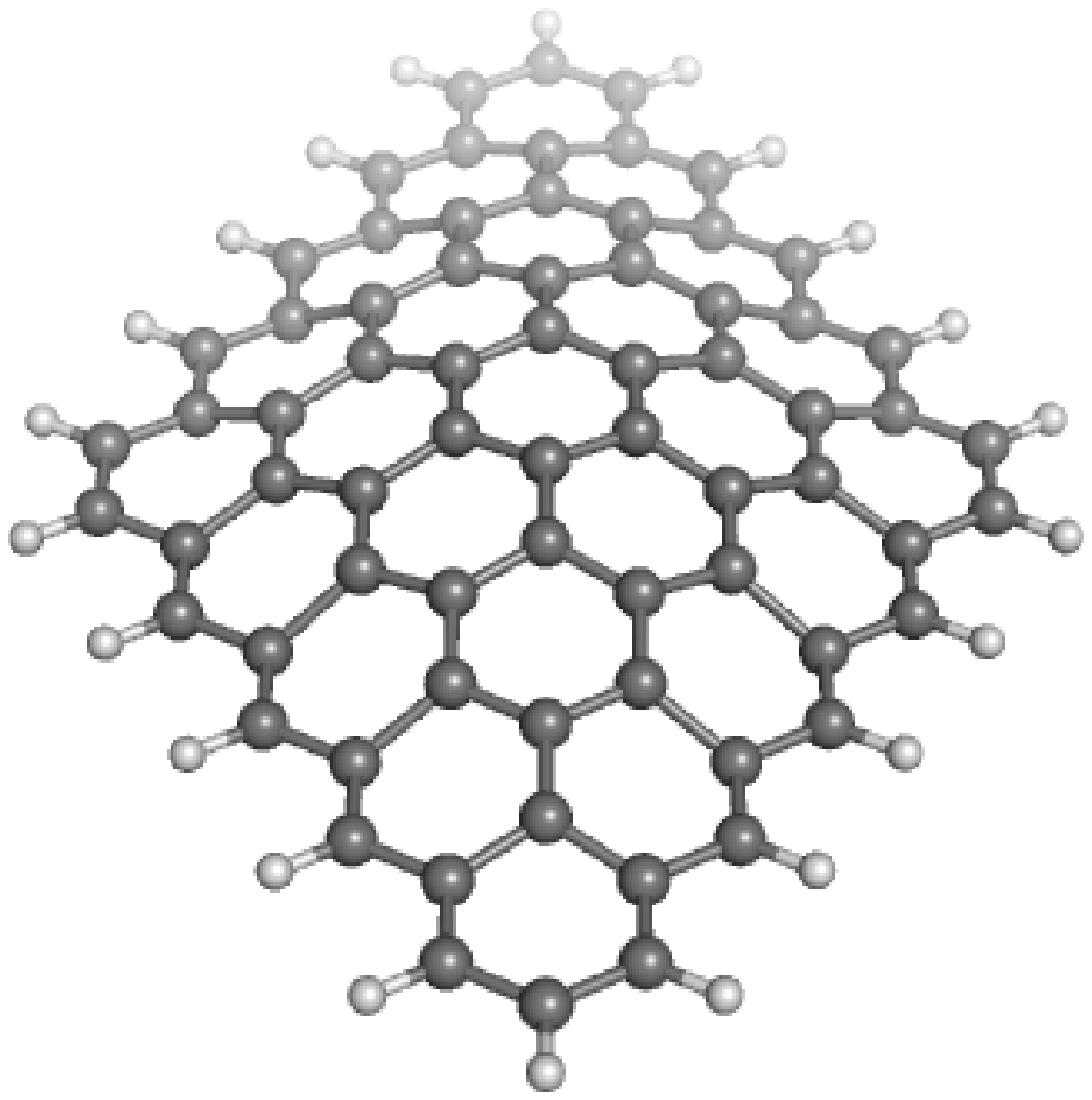}%
   \hspace*{0.3cm} \includegraphics[width=0.5\linewidth]{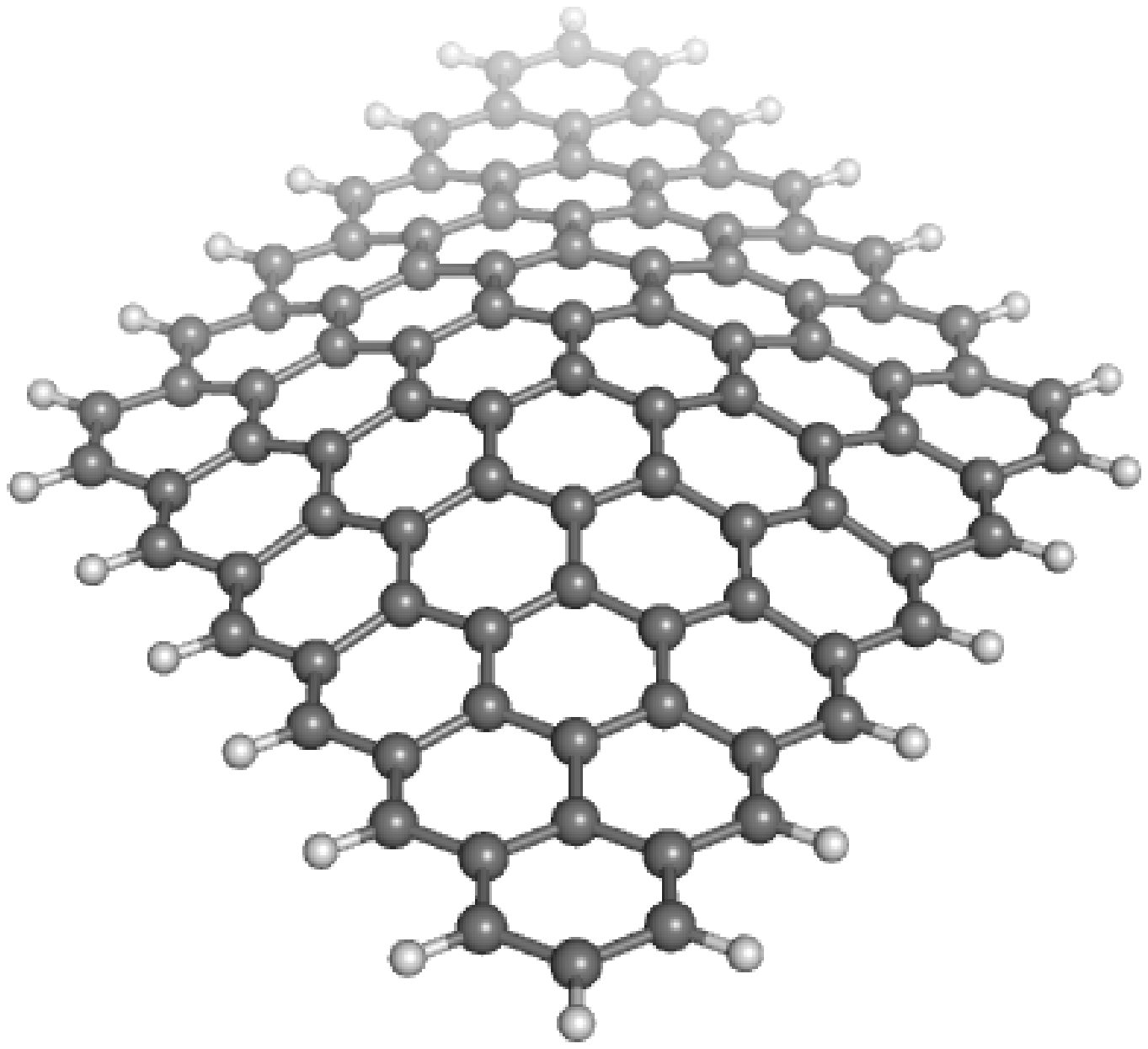}%
    \caption{Buckling flakes of $N$=5, 6 with different central
      configurations of carbon atoms.
      The atomic configuration of C atoms is relaxed under the
      constraint that the center atoms remain at a given height
    $h_0$ above the ground plane, while edge atoms (H and C) remain
    sitting within this plane ($h=0$).}
\label{f8}
\end{center}
\end{figure}

We present results from an additional DFT study, where we investigate the
transverse stiffness of the graphene flake that gives rise to the
elastic parameter $\kappa$. To this end we employ the following
strategy. Each flake has a center pair or center ring
of carbon atoms, see Figs.~(\ref{f1},\ref{f8}).
To create a transverse probing field $h({\bf r})$, we lift the center
atoms by the distance $h_0$ over the reference plane. After this, the
atomic structure of the flake is relaxed under the constraint, that the
set of {\it edge atoms} (H-atoms and edge C-atoms) can move only within the
reference plane; edge atoms cannot shift in $h$-direction.
\footnote{The motion of edge atoms in the plane is further restricted
  in the sense that only conformations have been considered, which can
be obtained by a homogenous rescaling of all C-C edge bond lengths.}
In this way, a flake is equipped with a single ripple while at the
same time the associated strain field ${\bf u}({\bf x})$ remains
negligibly small. In order to estimate the integrated curvature we numerically 
compute the bi-variate function which interpolates the scattered data values($h(\br)$-field)
at any predefined smooth mesh. We then use this interpolated function
to perform the second order numerical derivative at any arbitrary
precision.   

\begin{figure}[h]
\begin{center}
  \includegraphics[width=0.9\linewidth]{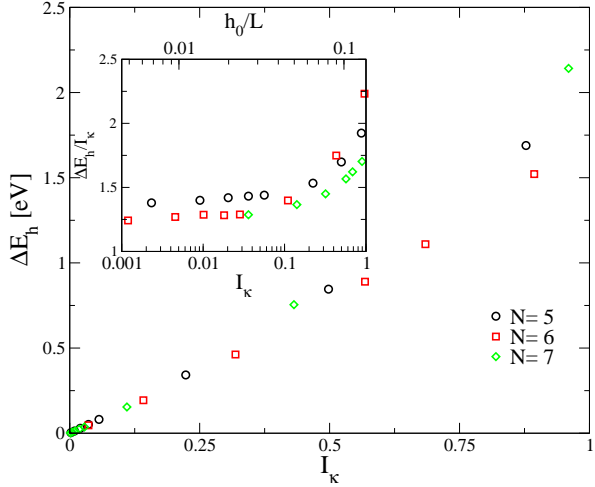}%
  \caption{Estimating the Lam\'e parameter $\kappa$ determined from the excess energy $\Delta E$ of a bulging flake with maximum height at $h_0$ over the unperturbed (flat) plane (see Fig.~\ref{f8}).
    Main panel: Change of energy $\Delta E$ with the
    integrated curvature $I_\kappa = \int d^2\br\ (\Delta h({\bf
      r}))^2$ for different flake sizes $N$.
    Inset: The ratio $\kappa=\Delta E/I_\kappa$ depend on $N$
    due to the effect of edge compression. }
  \label{f9}
\end{center}
\end{figure}

Fig.~\ref{f9} displays how the excess energy $\Delta E_{h}$
associated with the ripple grows with the increased integrated curvature,
\be
\label{e13}
I_{\kappa}= \frac{1}{2}\int \ d^2\br \ (\Delta h({\bf r}))^2 
\ee

The increase is linear, as expected from Eq.~(\ref{e2}) with a slope
that is only weakly dependent on the flake size, see inset
Fig.~\ref{f9}, this implies that 
nonlinearities remain small as long as 
the ratio of the ripples amplitude and wave length
, $h_0/L$, does not exceed $\sim$ 5\%.
The bending rigidity thus found is
$\kappa_{\rm{el}}$ = 1.24 eV which is well consistent
with the value 1.1eV obtained by
Fasolino, Los and Katsnelson\cite{Fasolino07}. 

Notice, that there is a significant scattering of almost 20\%
in earlier theoretical estimates for $\kappa$ and derived quantities,
see Tab. 2 in Ref. \onlinecite{Odegard02}. Discrepancies appear
because different theoretical techniques are being employed,
e.g. empirical potentials\cite{arroyo09} and density functional theory, 
but also because of modeling artefacts.
For example, extracting $\kappa$ from the elastic energy
of carbon nano-tubes (radius $R$) requires a very careful
extrapolation in $1/R$. If sub-leading terms are ignored, 
there is a pronounced tendency to overestimation, e.g. $\kappa=1.46$
eV in Ref. \onlinecite{kudin01}. These authors used nano-tubes with smaller
tube radius, hence bigger curvature, where
nonlinear effects become important. We can check the bending rigidity
 using the same curvature in Fig. \ref{f9} (inset) as reported
in Ref. \onlinecite{kudin01} and find a reasonable agreement with their value.



\section{Zero point motion}
In this chapter we extend our analysis of flake elastic properties and
take also the zero point motion of the atom cores into account,
that constitute the hexagonal lattice. Now, the free energy acquires a
second term,
\be
\label{e14}
F(N,d) = F_{\rm el}(N,d) + F_{\rm vib}(N,d)
\ee
with
\be
\label{e15}
F_{\rm vib}=\frac{1}{2} \sum_{p} \hbar \omega_p(N,d)
\ee
where $p$ labels all the flake's vibrational modes. The vibrational
excess energy associated with stretching the flake reads 
\be
\label{e16}
\Delta F_{\rm vib}=\frac{\hbar}{2} \sum_{p} \omega_p(N,d)-\omega_p(N,d(N))
\ee
where $\omega(N,d(N))$ denotes the vibration energies in the
absence of strain and $d(N)$ the equilibrium bond length, see inset 
Fig.~\ref{f5}.
Also $\Delta F_{\rm vib}$ can be expanded in terms
of the slow elastic modes,
\be
     \Delta F_{\rm{vib}}[h,{\bf
            u}]=\f{1}{2}\int_{{\cal A}}d^2{\bf r}\,
      \gamma_h\left(\Delta
      h({\bf r})\right)^2-\int_{{\cal
          A}}d^2{\bf
        r}\,\sum_{ij}\gamma_u^{ij}
      u_{ij}({\bf r}),
\label{e17}
\ee
with expansion parameters $\gamma_{h},\gamma_{u}$ that
represent averages of Gr\"uneisen parameters over 
all vibrational modes. In a two-dimensional sample\footnote{Recall that due to 
the presence of edge the symmetry group of our sample is lower ($D_{2h}$) than the 
one of the unperturbed hexagonal unit cell ($D_{6h}$).}
with a mirror symmetry
one expects $\gamma_u^{xy}=\gamma_u^{yx}=0$;
the change in phonon
frequencies should be even in the shear strain, $\varepsilon_{\rm s}$. 
Combining Eq.~(\ref{e17}) with an expansion of $\Delta F_{\rm el}$ in
full analogy with Eq.~(\ref{e2}) and after completing the square,
we find,
\begin{widetext}
       \be
       \Delta F_{\rm el \ vib}(N,d)=\f{\kappa_{\rm el}+ \gamma_h}{2} \ 
          \int_{\it A}d^2{\bf r} (\nabla^2 h)^2 +
          \f{\mu_{\rm el}+\lambda_{\rm
              el}}{2}\int_{A}d^2{\bf r}\Le(u_{ii}-\f{\gamma_{
                u}^{ii}}{\mu_{\rm el}+\lambda_{\rm
                el}}\Ri)^2 
      +\f{\mu_{\rm
            el}}{2}\int_{A}d^2{\bf r}\Le(u_{xx}-u_{yy}\Ri)^2 + u_{xy}^2  .
      \label{e18}
      \ee
\end{widetext}
For clarity, we have indicated in this expression
the bare electronic coefficients (i.e. with frozen atomic cores)
by $\kappa_{\rm el},\mu_{\rm el},\lambda_{\rm el}$. Likewise, the
displacement field ${\bf u}({\bf x})$ is defined with respect to the
optimum flake geometry ignoring vibrational terms.

In this way we can observe two facts.
(i) Vibrations modify the bare transverse
stiffness $\kappa_{\rm el}$ in Eq.~(\ref{e18}), 
$\kappa=\kappa_{\rm{el}}+\gamma_{h}$. 
(ii) Vibrations also affect interatomic distances. The effect can be
understood as an effective strain, which stretches the bare C-C
distances: $\gamma_{u}^{ii}/(\mu_{\rm el}+\lambda_{\rm el})$. 
(iii) The change in the C-C bond lengths eventually
feeds back into all elastic coefficients. Therefore, in a more
complete treatment of higher order terms also modifications in
$\mu_{\rm el}, \lambda_{\rm el}$ would occur. 
\subsection{DFT calculations of the Gr\"uneisen parameters}
In order to estimate the Gr\"uneisen parameters,
$\gamma_{h},\gamma_{u}$,
we should calculate the vibrational spectrum of flakes
with and without applied strain. To this end we adopt the following procedure.
For every flake size, $N$, we find the atomic geometry
with the optimal electronic energy, see e.g. Fig.~\ref{f1}.
This constitutes the set of freely relaxed ``parent states''. 
The relaxation ensures the Hessian, that characterizes interatomic
forces, to become a positive definite matrix. 
\footnote{This relaxation process is  the limiting computational step. For
a flake with $N=6$ it took several days on a single 
Opteron processor.  For $N{=}6$ a single
iteration takes about 1 day and for full convergence one
needs typically 50 days.}
\begin{figure}[t]
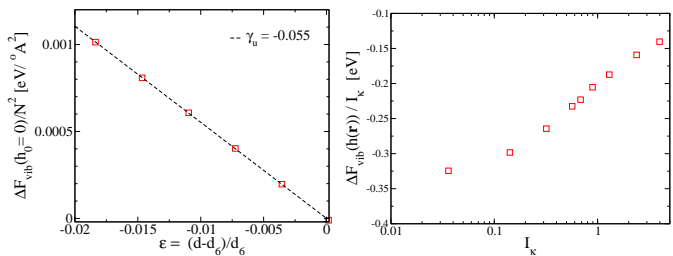

\begin{center}
   \includegraphics[width=0.5\linewidth]{f10a.g_u.eps}%
   \hspace*{0.1cm} \includegraphics[width=0.5\linewidth]{f10b.g_h.eps}%
   \caption{Left: Dependency of $\Delta F_{vib}$ on isotropic strain
     for system size $N{=}6$.  The slope defines $\gamma_u$ as in Eq.~(\ref{e17}).
     Right: Dependency of $\Delta F_{vib}$ on integrated curvature, $I_{\kappa}{=}\int \ d^2\br \ (\Delta h(\br))^2$, as in Fig.~\ref{f9}.} 
\label{f10}
\end{center}
\end{figure} 

In the present study we focus on the impact of phonons on the bulk
elastic constants. There we may eliminate
contributions of surface vibrations by assigning an infinite mass to
the surface H and C atoms. Other than this, the calculation of
vibrational modes and frequencies for the relaxed flake is a
standard procedure\cite{phonTB1,phonTB2}. 

Thereafter, each parent state thus obtained is used in order to create two new
families. The first family is constructed to obtain $\gamma_{u}$.
It derives by changing the bond length of edge
C-C-pairs by a factor of $1+\varepsilon$ keeping all atoms still
inside the base plane ($h{=}0$).
For each value $\varepsilon$ the internal C-atoms are relaxed
and the vibrational spectrum together with the average strain field,
$I_{u}(\varepsilon)=\int d{\bf r} (u_{xx}{+}u_{yy})$, 
are recalculated. In this process it is important to have edge atoms
at infinite mass. This ensures that the flake energy is in a
(constrained) minimum so that all frequencies are real.

In order to determine $\gamma_{h}$ a second family 
has been constructed. It consists of the buckled flakes,
Fig.~\ref{f8}, that we have
studied in the previous section in order to extract $\kappa_{\rm el}$. 
Again, after assigning infinite mass to the edge atoms for each family member, 
the vibrational spectrum and the consecutive modification of the
zero point energy can be calculated. 

\subsection{Results and Discussion}
\begin{figure}
\begin{center}
  \includegraphics[width=0.5\textwidth]{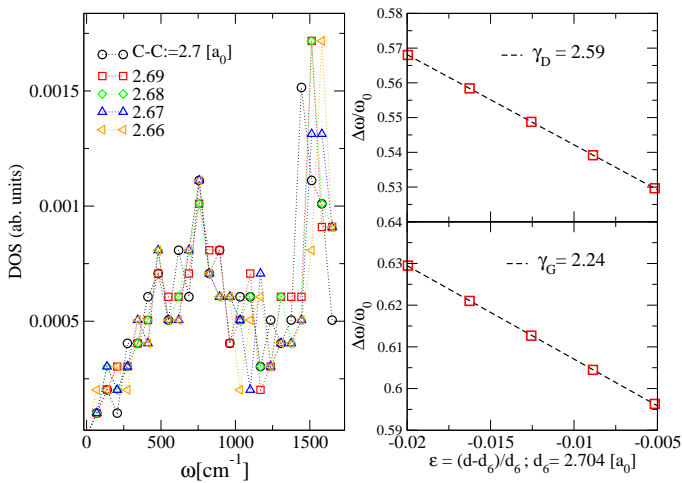}%
   \caption{Left panel: Phonon density of states for the flake of size $N{=}6$.
     Data sets are shown for different C-C bond distances. 
     Upper right panel shows the flow of D-peak
     with in-plane strain. Lower right panel shows the 
     variation of Raman G-peak with strain,
     where $\omega_0$ is the Raman band frequency($\approx$ 2680
     ${\rm cm}^{-1}$)\cite{Shen08}. 
    The corresponding vibration modes are shown in Fig.~\ref{f12}.}
\label{f11}
\end{center}
\end{figure}

\begin{figure}[th]
  \begin{center}
    \includegraphics[width=0.25\linewidth]{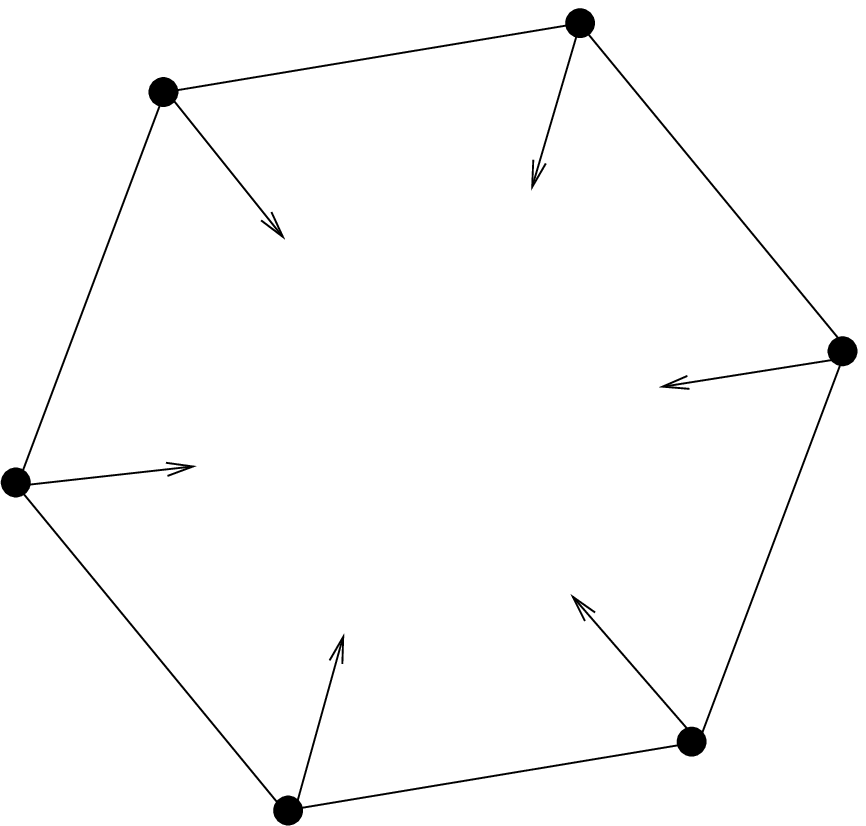}%
    \hspace*{0.6cm} \includegraphics[width=0.3\linewidth]{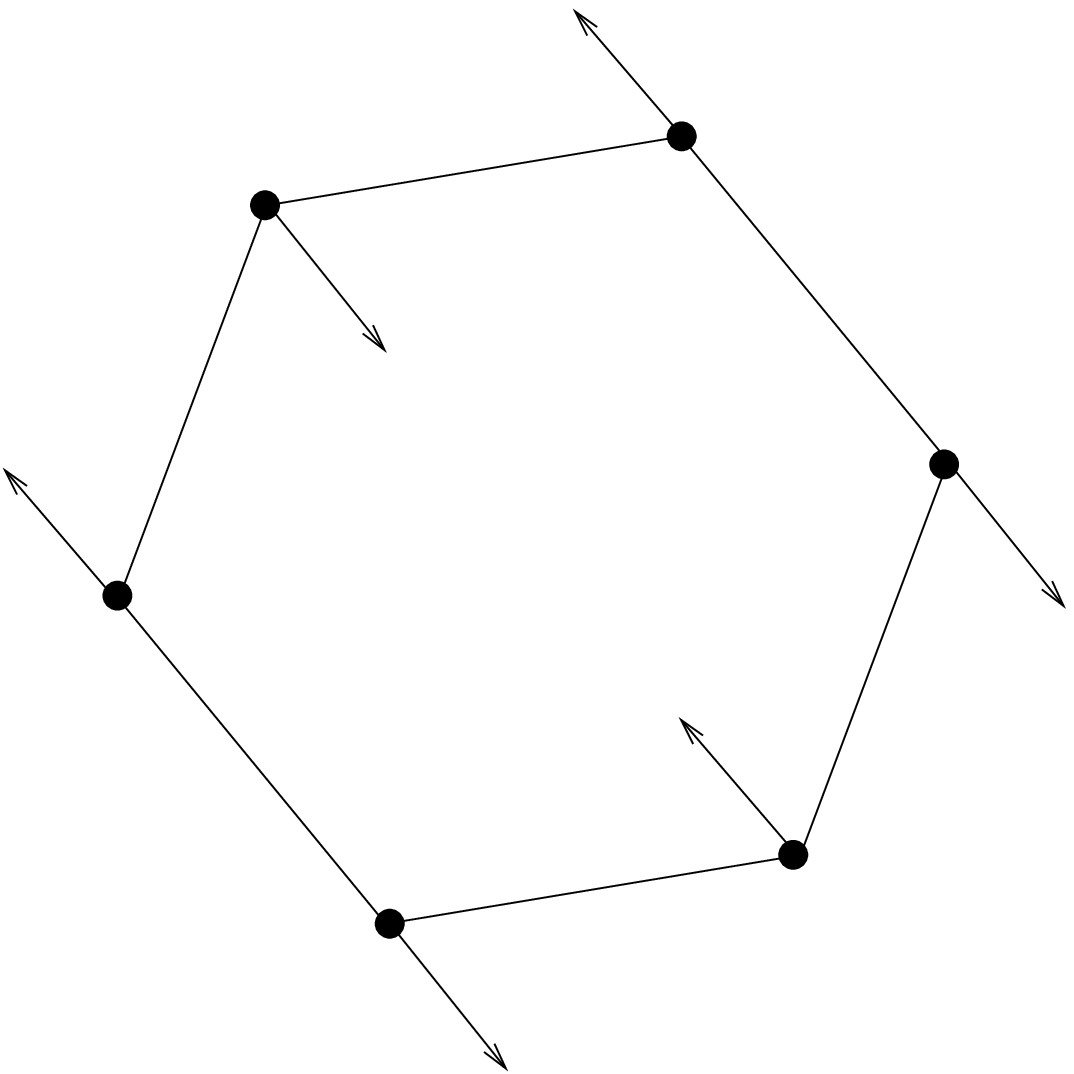}%
  \caption{Schematic diagram of vibrating modes that give rise to D(left) and 
    G(right) peaks in Raman spectra of graphene.}
  \label{f12}
  \end{center}
\end{figure}

In Figs.~\ref{f10} the change in the zero point energy,
$\Delta F_{\rm{vib}}$, is plotted over the integrated strain fields.
The Gr\"uneisen parameters are given by the slope near zero strain;
their numerical values are listed in Tab. \ref{t6}. 
\begin{table}
\begin{tabular}{c|c|c|c}
  &  BP86, $N{=}6$ & prev. calc. & expt. \\\hline
  $\gamma_{u}$ [eV/$\AA^2$] & -0.055 & - & - \\ 
  $\gamma_{h}$ [eV] &  -0.32 & - & - \\
  $\gamma_{u}/(\mu_{\rm el}+\lambda_{\rm el})$   & -0.004 & - & - \\ \hline
  $\gamma_{\rm D}$ [eV/$\AA^2$] & 2.6 & 2.7[\onlinecite{Marzari05}]& \\
  $\gamma_{\rm G}$ [eV/$\AA^2$]  & 2.2 & 2.0[\onlinecite{Thomsen02}] & $1.99$[\onlinecite{Mohiuddin09,proctor09}] \\
\end{tabular}
  \caption{Survey over the fitted Gr\"uneisen parameters extracted from the data
    Fig.~\ref{f10} and Fig.~\ref{f11} respectively.}
\label{t6}
\end{table}
For a discussion of our results we first recall, that
the vibrational spectral density of states of the carbon sheet
has a strong peak in the optical frequency regime, c.f. Fig.~\ref{f11}, 
near $1600 \ {\rm cm}^{-1}$. It is the ``G-peak'', that 
corresponds to an in-plane mode, where neighboring atoms vibrate
in opposite direction as depicted in Fig.~\ref{f12} (right).  
This mode gives the dominating
contribution to the total zero point energy.
Another significant contribution comes from the 
frequency range $500-1000 \ {\rm cm}^{-1}$, where one observes the 
mixing of out-of-plane modes with in-plane modes.  
A third important mode is the 
 ``D-peak'' near $1350 \ {\rm cm}^{-1}$,
that reflects the breathing mode as shown in Fig.~\ref{f12} (left). 
This mode is particularly
interesting when studying finite size(edge) effects in graphene flakes. The 
reason is that, in bulk graphene the coupling of the D-peak  to the electromagnetic fields
is suppressed, since the $D_{6h}$-symmetry  of the hexagonal unit cell remains
intact and inhibits the formation of the dipole moment. By
contrast, in flake with an overall symmetry that is lower than $D_{6h}$, the 
D-peak is observable with a strength proportional to the
inverse flake size.

Therefore, we can understand the sign of $\gamma_{h}$ as a consequence of a softening of these modes in the sample regions with non-zero curvature $(\Delta h({\bf r}))^2$. Similarly, $\gamma_{u}$ is negative, indicating the increase of the atomic oscillator frequency that occurs when the interatomic distance is diminished. 

While the sign of $\gamma_{h,u}$ was not unexpected, it is
noteworthy that the vibrational contributions to the phenomenological
material parameters are actually not so small.
The bare electronic bending rigidity,
$\kappa_{\rm el}$ is reduced by as much as 26\% down to
$\kappa{=}\kappa_{\rm el}+\gamma_{h}{=}0.88$eV. Similarly,
when expressing the effect of vibrations on the atomic lattice as an
effective strain pushing the atoms to larger distances, then this
strain reaches values up to 0.4\%.

Here, we also calculate the Gr\"uneisen parameters associated 
with individual modes (see Fig.~\ref{f12}). The right panel
in Fig.~\ref{f11} shows the flow of the Raman frequencies (upper half D, lower G) 
with the applied strain. The frequency decreases linearly 
 (as described in Eq.~(\ref{e17}))
with decreasing compression due to anharmonicity of the interatomic
potential. 
The slope essentially  estimates 
the Gr\"uneisen parameters, $\gamma_{\rm G}$ and $\gamma_{\rm D}$ for the vibrations
shown in Fig.~\ref{f12}. 
 Our results for $\gamma_{\rm G}$, $\gamma_{\rm D}$
are consistent with the earlier experiments and first
principal calculations (see Tab. \ref{t6}) available in literature\cite{Marzari05,Thomsen02,proctor09,Mohiuddin09}. 

\section{Conclusion}
The elastic properties of edge hydrogenated graphene flakes have been
investigated employing the density functional theory (DFT). 
Our study emphasizes the interplay between the edge
and bulk properties which are mediated via long range elastic forces. 

Specifically, we are able to disentangle bulk, surface and corner contributions to the
free energy together with the leading higher order corrections.
The binding energy per surface (edge) bond (7.5eV) is roughly two eV
higher than the one for interior (bulk) bonds (5.2eV); similarly,
edge bonds have a tendency to be shorter than bulk ones. As a
consequence, the flake's interior undergoes a surface induced
compression which is the more pronounced the smaller the flake 
is. This compression manifests itself in the way in which various
observables depend on the flake size, $N$.
For example, elastic constants (i.e. Lam\'e parameters)
of small flakes can exceed their bulk limit
($\mu+\lambda\approx 70$eV per ring, $\mu\approx 51$eV per ring,
$\nu=0.162$) by 30\% ($\mu+\lambda$) or even by 70\% ($\mu$).  
In comparison, the sheet (out of plane, buckling) stiffness, $\kappa\approx
1.2$eV, is less sensitive to $N$. Non-linearities remain weak (less
than 10\% increase) as long as the ratio of out of plane amplitude
and in plane wavelength of buckling is below 5\%. 
To highlight the importance of quantum effects on
elasticity we have also calculated the vibrational spectrum of
graphene flakes. Quantum corrections
affect mostly the sheet stiffness, $\kappa$,
lowering it significantly, about 26\% within our DFT framework. 

Finally, based on these results
we predict a pronounced shift of the Raman G- and D-peaks
with decreasing flake size to higher values. It
is a natural consequence of the edge induced flake compression.
The associated Gr\"uneisen parameters are
$\gamma_{\rm D}\approx 2.6$eV/$\AA^2$
and $\gamma_{\rm G}=2.2$eV/$\AA^2$.

 
\section{Acknowledgments}
We are grateful to  A. Bagrets, M. van Setten and V. Meded  for
helping us with the data visualization and also with the ab-initio
calculation. We are thankful to J. Weissm\"uller for numerous
discussions during this work. Also, we acknowledge support from the
Landesgraduiertenf\"orderung
Baden-W\"urttemberg and Center of Functional Nanostructures of
the Deutsche Forschungsgemeinschaft, project 4.11.

\bibliographystyle{apsrev}
\bibliography{lit}

\end{document}